\documentstyle{lamuphys}
\makeatletter
\let\chapter\hid@chapter
\makeatother
\begin{document}
\pagenumbering{arabic}
\title{Early Chemical Evolution of Galaxies}

\author{B.E.J. Pagel}

\institute{Astronomy Centre, CPES, University of Sussex, Brighton BN1 9QJ, UK}  

\maketitle

\begin{abstract}
Initial conditions are set by Big bang nucleosynthesis from which we know 
that 90 per cent of baryons are dark and have essentially unknown chemical 
composition. In our own Galaxy, there are many clues from individual stars in 
different populations whereas in elliptical galaxies the data largely come 
from integrated spectra, but these raise problems enough like the Mg/Fe 
and G-dwarf problems. Irregular and blue compact galaxies display 
the primary--secondary transition in 
N/O; this in turn may be relevant to element ratios observed in damped 
Lyman-$\alpha$ systems at high red-shift, which offer rather little evidence 
for pure SNII synthesis such as is found in the Galactic halo stars. A recent 
estimate of past star formation rates as a function of red-shift is presented 
and the appropriateness of the conventional conversion factor of 42 from 
SFR to metal production is discussed. For any reasonable value of this conversion 
factor, it is clear that most of the metals existing at $z=2.5$ have yet to be 
detected.  
\end{abstract}
\section{Introduction}

Initial conditions set by Big Bang nucleosynthesis are $Y=.24,\;Z=0$ for 
helium (e.g. Pagel 2000) and heavy elements respectively, D/H $=4\times 
10^{-5}$ (Levshakov, Tytler \& Burles 1998) and $^ 7$Li/H $=1.7\times 10^{-10}$ 
(Bonifacio \& Molaro 1997). The D/H ratio is the best indication of the overall
density of baryons in the universe, which can be expressed as $0.03\leq \Omega_ B 
h_{70}^ 2\leq 0.04$, similar to the density of Lyman-$\alpha$ forest gas at 
red-shifts 2 to 3 (Rauch et al. 1998), whereas the mass in visible stars in 
galaxies is given by $\Omega_ * h_{70}\simeq 0.0035$ (Fukugita, Hogan \& 
Peebles 1998), i.e. only 1/10 as much. Thus 90 per cent of baryonic matter is 
unseen and of unknown chemical composition, although it is reasonable to speculate 
that most of it is still intergalactic gas with $Z$ now somewhere between 
$0.3Z_{\odot}$ (Mushotzky \& Loewenstein 1997) and $0.1Z_{\odot}$ or less 
(Cen \& Ostriker 1999).

\footnotetext{Invited review given at Ringberg Workshop: {\em Galaxies in the 
Young Universe II}, August 2--6, 1999. Hans Hippelein (ed.), Springer-Verlag.} 

\input{psfig.sty} 

\begin{figure*}[htbp]   
\vspace{5cm} 
\includegraphics{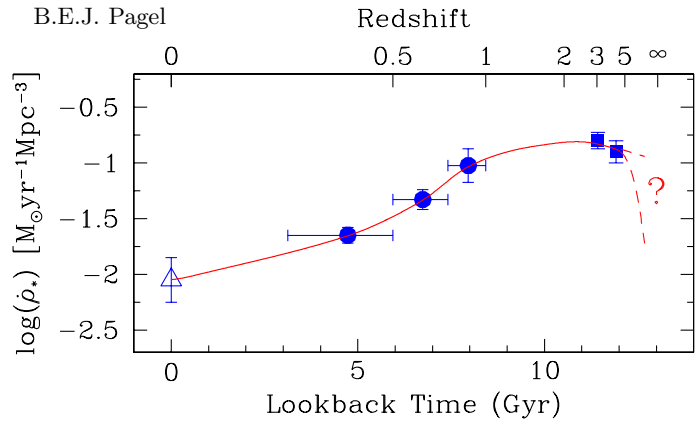}
\caption{Global comoving star formation rate density vs. lookback time 
compiled from wide-angle ground-based surveys (Steidel et al. 1999 and 
references therein) assuming E--de S cosmology with $h=0.5$, after 
Pettini (1999). Courtesy Max Pettini.}     
\label {fig1}
\end{figure*}  

%\begin{figure}
%\vspace*{-2.7cm} 
%\hspace*{-4.2cm} 
%\psfig{figure=pet99f5.ps, width=14cm,height=10cm,angle=270} 
%\vspace*{-4.3cm}  
%\caption{Global past rates of star formation after Pettini (1999). Courtesy 
%Max Pettini.} 
%\end{figure}  

The remainder of cosmic chemical evolution is the result of star formation, the 
history of which has been extensively studied by Madau and others (Madau et al. 
1996; Steidel et al. 1999; Pettini 1999) using data from red-shift surveys, 
Lyman break galaxies etc. 
(see Figure 1). Thus we now have a fair idea about global star formation rates 
since $z=4.5$, but ironically we do not know how to associate them with particular 
types of galaxies. The good news is that the integral over this version of the 
SFR history does come close to the estimated cosmic density of stars as given above, 
and it seems that about 1/4 of the stars were formed at red-shifts greater than 
2.5, over $10^{10}$ years ago. This raises the question of what happened to all the 
metals they made, to which I return in the last section.    

\section{Chemical Evolution of the Milky Way} 

\begin{figure*}[htbp] 
\vspace{5.5cm} 
\includegraphics{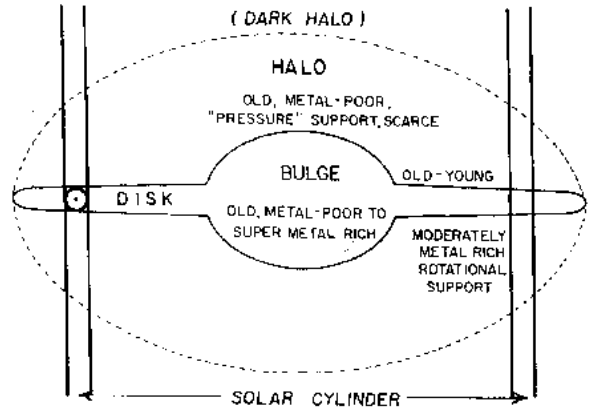} 
\caption{Schematic cross-section through the Galaxy.} 
\label{fig2}  
\end{figure*} 

The Galactic halo, bulge and disk(s) are all relevant to early times, only the thin 
disk being younger than the bulge and thick disk. The respective roles of 
hierarchical clustering, mergers and monolithic collapse are still not very clear;
probably all play a role, but the halo and bulge share a low specific angular 
momentum while the thick and thin disks share a high one and may result from 
later accretion of gas by the bulge, which would then resemble an E-galaxy. 
However, it is also possible that the bulge evolved from the disk by way of a bar. 

In any case, the stellar dynamics of the halo favour what Thomas, Greggio \& Bender 
(1999) refer to as a `fast clumpy collapse', basically the old idea of Eggen, 
Lynden-Bell \& Sandage (1962) placed in the context of modern hierarchical clustering 
scenarios. One point of interest is the metallicity distribution function (MDF), 
recently extended to very low metallicities by Beers et al. (1998). The MDF is 
essentially the modified Simple-model type distribution originally noted by 
Hartwick (1976), with a peak at about 1/10 of the true yield, down to [Fe/H] 
$\simeq -3$. Below that it begins to fall off and there are virtually no stars 
(compared to a predicted number of about 10) below $-4$, which could represent  
enrichment either from a hypothetical Population III or from  contamination of 
low-mass stars by a nearby supernova. 
   
\begin{figure*}[htbp] 
\vspace*{8.5cm} 
\includegraphics{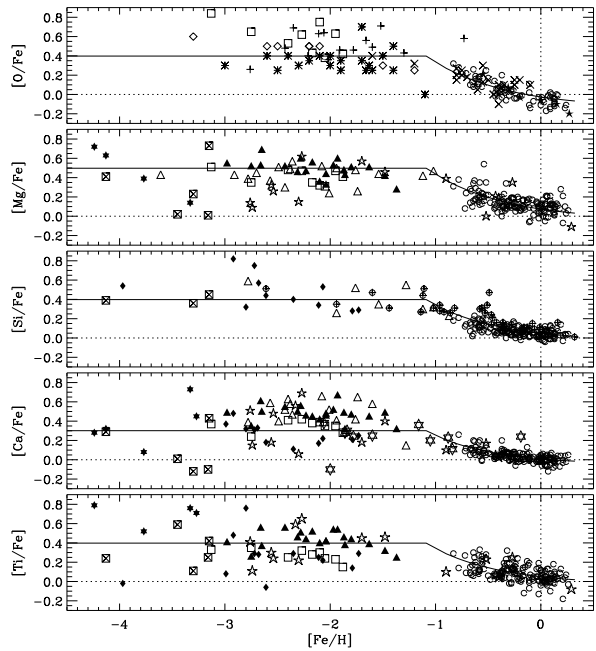} 
\caption{Abundance ratios of oxygen and $\alpha$-elements to iron, plotted against 
[Fe/H] for stellar samples from the Galactic disk and halo, after Pagel \& 
Tautvai\v sien\. e (1995). The solid line and curve represent a simple analytical 
Galactic chemical evolution model.} 
\label{fig3}  
\end{figure*} 

A significant clue to early Galactic chemical evolution comes from the relation 
between oxygen and $\alpha$-particle elements, thought to come exclusively or mainly 
from type II supernovae, and iron, more than half of which in the Solar System comes 
from Type Ia. Fig 3 suggests that there is a plateau in O,$\alpha$/Fe at low 
metallicities (assumed to represent early times), but there is currently a 
controversy in the case of oxygen.  Abundances derived from the forbidden [OI] line, 
which is probably the most reliable source when it is not too weak, suggest a 
plateau, but from measurements of the near UV OH bands in dwarfs and subgiants, 
both Israelian, Garc\' ia Lopez \& Rebolo (1998) and Boesgaard et al. (1999) 
have derived a rising trend with diminishing [Fe/H] more or less following the 
open squares in the top panel of Figure 3. In contrast, Fulbright \& Kraft (1999) have 
studied the [OI] spectral region in two of the extreme cases and find lower O/Fe 
ratios fitting the plateau.  There are technical difficulties in both methods:
the OH bands are subject to uncertainties in UV continuum absorption (cf. 
Balachandran \& Bell 1998 on solar beryllium abundance) and effective temperature, 
while the forbidden line in the relevant cases is so very weak that the definition 
of the continuum becomes a crucial source of uncertainty.        
 
However this controversy comes out, the O,$\alpha$ enhancement is not universal,  
as has been shown, e.g. by  Nissen \& Schuster (1997); there are `anomalous' halo 
stars which have more solar-like element ratios even at quite low metallicities, 
a feature that is also found in the Magellanic Clouds and can be explained on the 
basis of slower star formation rates and effective yields diminished by outflows 
(e.g. Pagel \& Tautvai\v sien\. e 1998). However, within the halo the presence 
of `anomalies' shows no obvious relation with extreme kinematic properties that might 
be signatures of a captured satellite (Stephens 1999). 

\begin{figure*}[htbp] 
\vspace{6cm}
\includegraphics{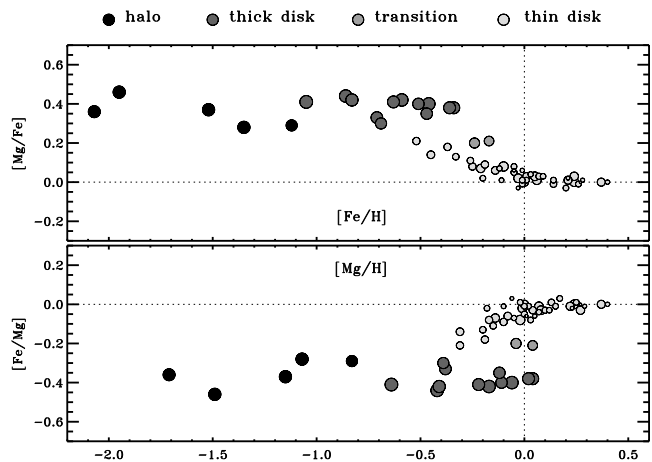} 
\caption{[Mg/Fe] vs [Fe/H] and [Fe/Mg] vs [Mg/H] for stars of the Galactic halo, 
thick disk and thin disk, after Fuhrmann (1998). Courtesy Klaus Fuhrmann.}   
\label{fig4} 
\end{figure*} 

Within the thick disk, the $\alpha$/Fe ratio is remarkably uniform, even up to 
quite high metallicities, indicating an old `get rich quick' population. This 
is well brought out by the work of Fuhrmann (1998) on Mg, shown in Figure 4, and 
in a still unpublished study of oxygen by Gratton et al. (1996), and it may be that 
this trend is continued in the bulge (cf. Rich 1999). The data cast an interesting 
light on the formation of the thick disk, since they indicate a hiatus in star 
formation during which Fe/$\alpha$ increased but overall metallicity diminished, 
maybe from inflow of relatively unprocessed material, e.g. in a merger, before the
stars now belonging to the thin disk were formed.   

Returning to the earliest stage of evolution of the Population II halo, when we 
consider a regime in which [Fe/H] $<-2.5$ or so, we reach a stage where 
pollution by a single supernova becomes significant over a region the size of 
a globular cluster or superbubble of the order of $10^ 5 M_{\odot}$. Metallicity 
(however defined) then becomes a poor clock and strange patterns appear, 
accompanied by significant scatter (McWilliam 1997). There are marked changes 
within the iron group, with Cr, Mn (and Cu) going down relative to iron and Co 
going up.  Ryan, Norris \& Beers (1996) suggest that at these low levels [Fe/H] 
is an increasing function of the mass of an individual supernova, and Tsujimoto 
\& Shigeyama (1998) have estimated revised stellar yields as a function of progenitor 
mass on this basis. Most yields increase, with the conspicuous exception of the 
r-process, whose representative Eu/Fe has a large scatter and may be anti-correlated 
with [Fe/H]. Ba and Sr also mainly come from the r-process at these low metallicities 
and have even more scatter because the s-process can also contribute in evolved 
stars or stars with evolved companions. In a model recently put forward by 
Tsujimoto, Shigeyama \& Yoshii (1999), stars form in superbubbles dominated by a 
single supernova, so that their composition is a weighted mean of the interstellar 
medium (with [Eu/Fe] $\simeq [\alpha$/Fe] = constant)  and supernova ejecta.  Fe/H 
increases with the mass of the supernova while Eu/Fe decreases, leading to an 
anti-correlation with scatter superimposed until the ISM is sufficiently enriched to 
take over and normal Galactic chemical evolution proceeds.    

Further evidence for inhomogeneity comes from the abundances of the light elements 
$^ 6$Li, beryllium and boron, which show an unexpected `primary' behaviour --- at 
least relative to iron --- down to 
very low metallicities. This cannot be understood on the basis of spallation of 
interstellar CNO nuclei by primary cosmic ray protons and $\alpha$-particles; these  
give a reasonable explanation for their abundances in the Sun and Population I stars 
in general but led to an expectation of secondary behaviour (Be,B/O $\propto$ O/H) 
with diminishing metallicity. \footnote{With the large increase in O/Fe claimed by 
Israelian et al. and Boesgaard et al. there could be some semblance of secondary 
behaviour of the light elements after all, along with iron, magnesium, calcium etc; 
the likelihood of this depends on how the oxygen debate comes out.}  There are also 
energetic problems with the production by interstellar spallation at low metallicity 
(Ramaty et al. 1997). Thus various inhomogeneous processes have been proposed, 
beginning with the hypothesis of Duncan, Lambert \& Lemke (1992) that fast CNO 
nuclei in primary cosmic rays are reponsible, and that their abundance is dominated 
by supernova ejecta rather than the interstellar medium.  A more detailed model by
Ramaty \& Lingenfelter (1999) postulates an origin of of cosmic rays from acceleration 
of ions sputtered off dust grains in supernova ejecta by shocks within a superbubble. 
Thus the composition of cosmic rays is more or less constant and they dominate light 
element production at early times in the way suggested by Duncan, Lambert \& Lemke.   

\section{Elliptical Galaxies} 

Most of our information on E-galaxies comes from colours and spectral features of 
integrated light interpreted with the aid of population synthesis models based on 
the theory of stellar evolution and a spectral library.  A classical result is the 
correlation between the Lick Mg$_ 2$ index and central velocity dispersion 
(Bender 1992). For an old population, Mg$_ 2$ should be a good measure of the overall 
heavy-element abundance $Z$, dominated by oxygen and other $\alpha$-elements, 
because Mg itself is one of these and Mg and Si supply 2/3 of the free electrons 
providing H$^ -$ opacity in red-giant atmospheres. However, age is a complication 
and the correlation with iron is more problematic (cf. Figure 5).  At face value, 
based on single stellar population (SSP) models by Worthey (1994) and by Buzzoni 
(1995), the nuclear $Z$ or Mg abundance increases with depth of the potential 
well, whereas that of iron does not: the Mg/Fe dilemma. According to theoretical 
simulations by Thomas, Greggio \& Bender (1999) and Thomas \& Kauffmann (1999), 
the expectation would be that star formation goes on for longer in the bigger 
E-galaxies, making their weighted-mean age smaller and Mg/Fe smaller rather than 
larger.   

\begin{figure*}[htbp] 
\vspace{8cm} 
\includegraphics{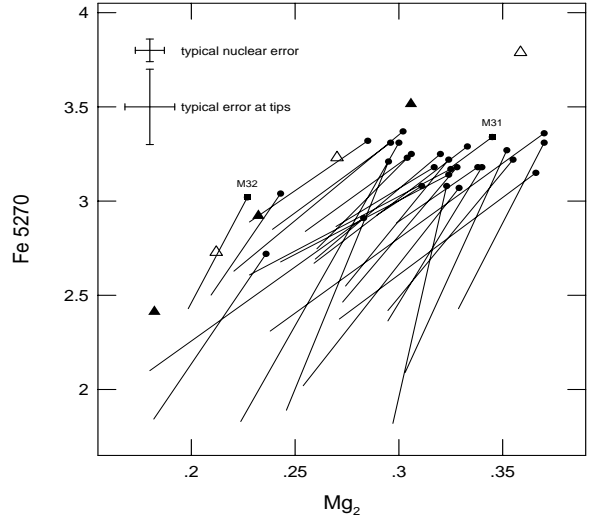} 
\caption{Plot of an iron feature against Mg$_ 2$. Filled circles and squares 
represent the nuclear regions (central 5 arcsec) of elliptical galaxies, while the 
sloping lines show the mean trend with galactocentric distance in each one. 
Triangles show model predictions for ages of 9 (solid) and 18 Gyr (open), based on 
SSP models that fit features in globular clusters assuming [Mg/Fe] = 0. A young 
model with [Fe/H] = 0 fits the nucleus of M 32 quite well, and the predicted trends 
with metallicity run roughly parallel to several of the observational lines, but 
the trend among nuclei is not fitted at all. After Worthey, Faber \& Gonzalez 
(1992). Courtesy Guy Worthey.} 
\label{fig5} 
\end{figure*} 

There is also a `G-dwarf' problem, at least for nuclei, in the sense that SSP 
models fit the UV spectra better than those incorporating simple models of galactic
chemical evolution (Bressan, Chiosi \& Fagotto 1994; Worthey, Dorman \& Jones 1996;
 Greggio 1997). 
One suggestion has been that the nuclei are pre-enriched with processed infalling 
material during a rapid  clumpy collapse (Greggio 1997).  The `concentration 
model' of Lynden-Bell (1975) may also be relevant to this situation, but according 
to Worthey, Dorman \& Jones this is not just a nuclear problem.    

Some notable results emerge from the recent study by J\o rgensen (1999) of spectral 
features of galaxies in the Coma cluster. She confirms the existence of an 
age-metallicity relation as envisaged in the numerical 
simulations of Thomas \& Kauffmann (1999), both for iron and magnesium, 
consistent with the view that galaxies with deep enough potential wells to hold 
on to their gas for longer reach higher metallicities. At any age, the galaxies 
with the highest velocity dispersions have the highest metallicity judged from 
magnesium, but for iron quite anomalously the opposite is the case, which makes 
one wonder about the calibration. Finally, Mg/Fe is independent of age and 
increases with velocity dispersion, which is hard to explain on the basis 
of the orthodox view of the unaided effects of a time lag for SNIa. Thomas 
(1999) has suggested that galactic nuclei may be affected by sporadic starbursts with
a flat IMF. 

The question of the IMF, or at least the yield, is also raised by the supply of 
iron and other elements to the X-ray gas in rich clusters of galaxies. Adapting an 
argument due to Renzini et al. (1993), we can start from the empirical finding of 
Arnaud et al. (1992) that the total mass of iron in the gas is proportional to 
the total optical luminosity of the E and S0 galaxies in the cluster according to  
\begin{equation}
\frac{M_{{\rm Fe}}({\rm gas})}{L_ *}=\frac{1}{55}\,\frac{M_{\odot}}{L_{\odot}}, 
\end{equation}  
whence if $M_ */L_ *\leq 10$ solar units, then 
\begin{eqnarray} 
\frac{M_{\rm Fe}({\rm gas})}{M_ *}\geq 1.8\times 10^{-3} &=&1.5Z_{\odot}({\rm Fe})
\nonumber \\
 \frac{M_{\rm Fe}(*)}{M_ *} &\simeq& Z_{\odot}(\rm Fe) \nonumber \\
{\rm Overall\;true\; yield}   &=& 2.5Z_{\odot}, 
\end{eqnarray} 
where the overall true yield \footnote{Defined as the mass of newly synthesised and 
ejected heavy elements from a generation of stars divided by the mass remaining in 
long-lived stars and compact remnants (Searle \& Sargent 1972).} has been obtained 
by simply dividing the mass of iron by the 
mass of the stars; since the iron: oxygen ratio is about solar, the same result 
would have been obtained if we had considered oxygen instead of iron.  This yield, 
however, is very high in comparison with values of $Z_{\odot}$ or slightly less 
that come up in studies of chemical evolution in the solar neighbourhood (e.g. 
Pagel \& Tautvai\v sien\. e 1995), raising the question of whether such a high 
value is actually universal and the lesser yields found in other contexts just 
a consequence of mass loss from the systems. If so, it would be sufficient to 
enrich the intergalactic medium to the 1/3 of solar value postulated by 
Mushotsky \& Loewenstein (1997).  

\section{Abundances at High Red-shift}

\begin{figure*}[htbp] 
\vspace{6cm} 
\includegraphics{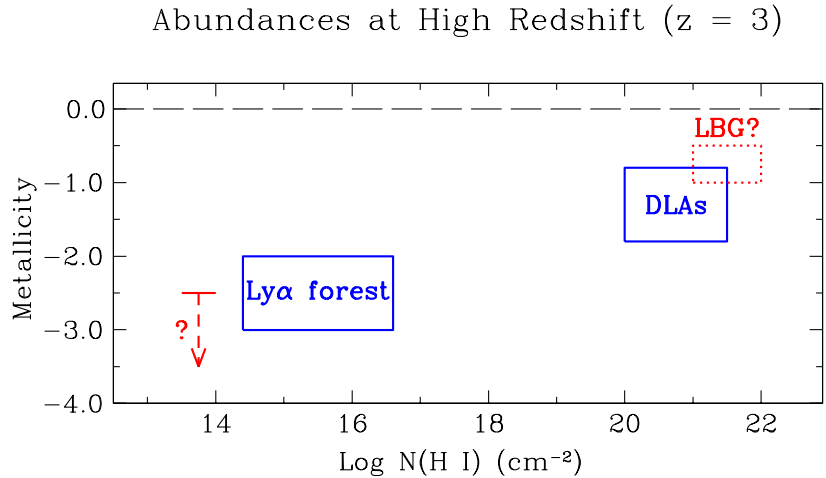} 
\caption{Summary of our current knowledge of abundances at high red-shift. 
Metallicity is on a log scale relative to solar and $N$(H I) is the column density 
of neutral hydrogen measured in the Lyman-$\alpha$ forest, damped Lyman-$\alpha$ 
systems and Lyman break galaxies, after Pettini (1999). Courtesy Max Pettini.} 
\label{fig6} 
\end{figure*} 

Naturally recent advances in studies of objects at high red-shift supply vital clues 
to the early evolution of galaxies, but, as Pettini (1999) has emphasised, our 
knowledge in this area is severely limited (see Figure 6), giving rise to serious  
observational selection effects. The Lyman forest comes from condensations in 
the intergalactic medium, possibly analogues of the high-velocity H I clouds seen 
today (Blitz et al. 1999), and represents the majority of the baryonic matter in 
the universe, while the damped Lyman-$\alpha$ (DLA) systems have a co-moving 
density similar to that of disk galaxies today.  Then there are also the Lyman 
break galaxies, for which there is some information based on the strength of their 
stellar winds. Figure 7 shows the metallicities of DLA systems, based on zinc 
abundance, plotted against red-shift, after Pettini (1999). When column-density 
weighted means are formed in distinct red-shift bins,  
no evolution is detectable in the metallicity and there 
is no obvious way of identifying what sort of objects these systems will eventually 
become. Some clues could come from element:element ratios like N/O or $\alpha$/Fe. 
Here the difficulty lies in correcting for depletion from the gas phase on to 
dust, which can be estimated (when not too large) from the ratio of Zn to Cr and 
Fe, since their intrinsic relative abundances are usually constant. According to 
Vladilo (1998) and Pettini et al. (1999ab), the resulting relative abundances of 
silicon and iron are pretty much solar (or like the Magellanic Clouds and the 
`anomalous' halo stars referred to above), suggesting that they are destined to 
become Im galaxies rather than large spirals. The behaviour of N/Si vs Si/H  also 
shows a resemblance to the behaviour of N/O vs. O/H in  
irregular and blue compact galaxies with perhaps an even greater  
scatter around the normal primary-secondary pattern than is found in irregulars 
and BCGs (Lu, Sargent \& Barlow 1998).

\begin{figure*}[htbp]
\vspace{5cm}
\includegraphics{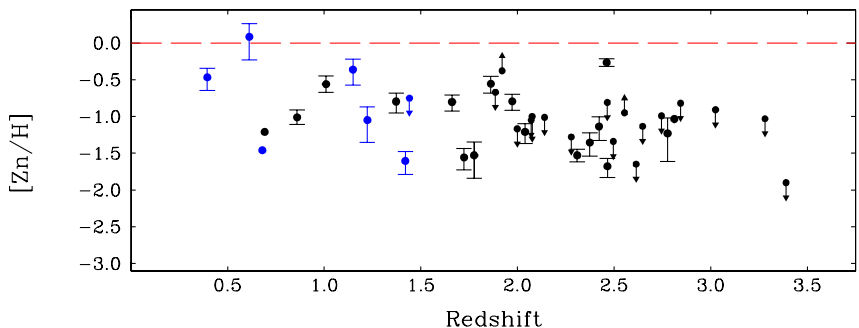}          
\caption{Zn abundance against red-shift for 40 DLAs from Pettini et al. (1999). 
Courtesy Max Pettini.} 
\label{fig7} 
\end{figure*} 

What are the consequences of the new star formation rate density (Fig 1) for `metal' 
production and global chemical evolution? To begin with, the SFR which I shall call 
$\dot{\rho}_ *$(conv.) is based on the rest-frame UV luminosity density combined with 
a Salpeter power-law IMF between 0.1 and $100M_{\odot}$. The co-moving metal 
production-rate density is then usually deduced by dividing by the magic number of 
42 (shades of {\em The Hitch-Hiker's Guide to the Galaxy}), which comes from 
models of supernova yields in the range of 10 to $100M_{\odot}$ or so, and I shall 
call this metal production rate $\dot{\rho}_ Z$(conv.). The overall yield then amounts 
to 
\begin{equation}
y=\frac{\dot{\rho}_ Z(\rm conv.)}{\alpha\dot{\rho}_ *(\rm conv.)}= 
\frac{1}{42\alpha}\simeq 0.03\simeq 2Z_{\odot}, 
\end{equation} 
where $\alpha\simeq 0.7$ is the lock-up fraction.  Such a high yield is excessive 
for the solar neighbourhood (although it may be suitable for intra-cluster gas) 
and so people modelling Galactic chemical evolution generally either use a steeper 
slope, a smaller lower mass limit or assume that stars above 40 or 50$M_{\odot}$ 
lock the bulk of their element production in black holes. So the true rate of 
`metal' production should be $\beta\dot{\rho}_ Z$(conv.), where $\beta\leq 1$ is 
some correction factor depending on your favourite model of galactic chemical 
evolution. Finally, the true star formation rate density should be corrected by 
some factor $\gamma$, also $\leq 1$, for the undoubted flattening of the IMF 
power law somewhere below $1M_{\odot}$, e.g. Fukugita, Hogan \& Peebles (1998) 
have $\gamma = 0.65$, but this does not influence the conversion factor (at least 
to first order) because it is mainly just the massive stars that produce both the 
metals and the UV luminosity.  

With these preliminaries, we can use the data supplied by Pettini (1999) to draw 
up the following inventory of stars and metals for the present epoch and for 
a red-shift of 2.5, assuming $\alpha =0.67$, $\gamma = 0.65$. 

\begin{center} 
\begin{table} 
{\large 
\caption{Inventory of stars and metals at $z=0$ and $z=2.5$} 
\label{tab1} 
\begin{tabular}{|r|c|c|} 
\hline
&$z=0$ & $z=2.5$ \\ \hline 
&&\\  
$\rho_ * = \alpha\,\gamma\int\dot{\rho_ *}{\rm (conv.)}\,dt$ & 
$3.6\times 10^ 8\,  
 M_{\odot}\,{\rm Mpc}^{-3}$ & $9\times 10^ 7\,M_{\odot}\, {\rm Mpc}^{-3}$\\ 
$\Omega_ *= \rho_ */7.7\times 10^{10}h_{50}^2$ & $.0047h_{50}^{-2}$ & 
$.0012h_{50}^{-2}$ \\ 
$\Omega_ *$(FHP 98) & $.0049h_{50}^{-1}$ &\\ 
$\rho_ Z=y\rho_ *=\beta\rho_ */(42\alpha\gamma)$ & $2.0\times 10^ 7\beta\,  
M_{\odot}\,{\rm Mpc}^{-3}$ & $5\times 10^ 6 \beta \,M_{\odot}\, 
{\rm Mpc}^{-3}$\\ 
$\Omega_{Z}$ (predicted) & $2.6\times 10^{-4}\beta h_{50}^{-2}$ & 
$6.5\times 10^{-5}\beta h_{50}^{-2}$\\ 
$\Omega_ Z$ (stars, $Z=Z_{\odot}$) & $1.0\times 10^{-4}h_{50}^{-1}$ &\\ 
$\Omega_ Z$ (hot gas, $Z=0.3Z_{\odot}$) & $1.7\times 10^{-4}h_{50}^{-1.5}$&\\
&$\Rightarrow 0.4\leq\beta\leq 1$&\\   
$\Omega_ Z$ (DLA, $Z=0.07Z_{\odot}$) & & $3\times 10^{-6}h_{50}^{-1}$\\ 
$\Omega_ Z$ (Ly. forest, $Z=0.003Z_{\odot}$) & & $4\times 10^{-6}h_{50}^{-2}$\\
$\Omega_ Z$ (Ly. break gals, $Z=0.3Z_{\odot}$) & & ? \\
$\Omega_ Z$ (hot gas) & & ?\\ \hline       

\end{tabular} } 
\end{table}
\end{center}

The $z=0$ column shows a fair degree of consistency.  We can live with $\beta =1$ 
if we wish to explain a metal content of intergalactic gas as high as suggested by 
Mushotzky \& Loewenstein, or we can take this as a firm upper limit because we do not
know if there is that much `metal' in intergalactic gas. 

Somewhat more troubling questions arise at red-shift 2.5, however, as Pettini 
(1999) has already pointed out.  It now seems that about a quarter of the stars have 
already been formed by then (in ellipticals, bulges and 
thick disks?), but known entries in the table only account for 10 per cent of the
resulting metals (if $\beta=1$) or 25 per cent (if $\beta=0.4$). This is a good 
measure of the incompleteness in our knowledge of the distribution of the elements 
at substantial red-shifts. 

I thank Max Pettini for supplying data and for enlightening discussions.

\section*{References} 
\hspace*{-9mm}
{\leftskip 8mm 
\parindent -8mm
Arnaud, M., Rothenflug, R., Boulade, O., Vigroux, L. \& Vangioni-Flam, E. 1992, 
AA, 254, 49 

\parskip 0mm

Balachandran, S.C. \& Bell, R.A. 1998, Nature, 392, 791   

Beers, T.C., Rossi, S., Norris, J.E., Ryan, S.G., Molaro, P. \& Rebolo, R. 
1998, in {\em Primordial Nuclei and their Galactic Evolution}, N. Prantzos, 
M. Tosi \& R. von Steiger (eds.), Sp. Sci. Rev., 84, 139 

Bender, R. 1992, in IAU Symp. no. 149, {\em Stellar Populations of Galaxies}, 
B. Barbuy \& A. Renzini (eds.), Kluwer, Dordrecht, p. 267 

Blitz, L. 1999, in IAC Euroconference: {\em The Evolution of Galaxies on 
Cosmological Timescales}, J.E. Beckman \& T.J. Mahoney (eds.), ASP Conf. 
Series, in press    
 
Boesgaard, A., King, J.R., Deliyannis, C.P. \& Vogt, S.S. 1999, AJ, 117, 492 
 
Bonifacio, P. \& Molaro, P. 1997, MNRAS, 285, 847  

Bressan, A., Chiosi, C. \& Fagotto, F. 1994, ApJS, 94, 63 
 
Buzzoni, A. 1995, ApJS, 98, 69  
 
Cen, R. \& Ostriker, J.P. 1999, ApJ, 514, 1 

Duncan,D., Lambert, D.L. \& Lemke, M. 1992, ApJ, 401, 584  

Eggen, O.J., Lynden-Bell, D. \& Sandage, A.R. 1962, ApJ, 136, 748 

Fuhrmann, K. 1998, AA, 338, 161 
 
Fukugita, M., Hogan, C.J. \& Peebles, P.J.E. 1998, ApJ, 503, 518

Fulbright, J.P. \& Kraft, R.P. 1999, AJ, 118, 527  

Gratton, R., Carretta, E., Matteucci, F. \& Sneden, C. 1996, preprint 

Greggio, L. 1997, MNRAS, 285, 151  

Hartwick, F.D.A. 1976, ApJ, 209, 418 

Israelian, G., Garc\' ia Lopez, R.J. \& Rebolo, R. 1998, ApJ, 507, 805 

J\o rgensen, I. 1999, MNRAS, 306, 607  
 
Levshakov, S., Tytler, D. \& Burles, S. 1998, astro-ph/9812114, AJ, subm.

Lu, L., Sargent, W.L.W. \& Barlow, T.A. 1998, AJ, 115, 55 

Lynden-Bell, D. 1975, Vistas in Astr, 19, 299 

McWilliam, A. 1997, ARAA, 35, 503  

Madau, P., Ferguson, H.C., Dickinson, M.E. et al. 1996, MNRAS, 283, 1388 

Mushotzky, R.F. \& Loewenstein, M. 1997, ApJ, 481, L63  

Nissen, P.E. \& Schuster, W.A. 1997, AA, 326, 751 
 
Pagel, B.E.J. 2000, Phys. Rep., in press 

Pagel, B.E.J. \& Tautvai\v sien\. e, G. 1995, MNRAS, 276, 505 

Pagel, B.E.J. \& Tautvai\v sien\. e, G. 1998, MNRAS, 299, 535 

Pettini, M. 1999, in ESO Workshop: {\em Chemical Evolution from Zero to High 
Redshift}, J. Walsh \& M. Rosa (eds.), Springer-Verlag, in press, 
astro-ph/9902173   

Pettini, M., Ellison, S.L., Steidel, C.C. \& Bowen, D.V. 1999a, ApJ, 510, 576

Pettini, M., Ellison, S.L., Steidel, C.C., Shapley, A.E. \& Bowen, D.V. 1999b, 
preprint astro-ph/9910131, ApJ, in press    

Ramaty, R., Kozlovsky, B., Lingenfelter, R. \& Reeves, H. 1997, ApJ, 488, 730  

Ramaty, R. \& Lingenfelter, 1999, in {\em LiBeB, Cosmic Rays, and Related X- and 
$\gamma$-Rays}, R. Ramaty, E. Vangioni-Flam, M. Cass\' e \& K. Olive (eds.), 
ASP Conf. Series, Vol. 171, p. 104  

Rauch, M., Miralda-Escud\' e, J., Sargent, W.L.W. et al. 1998, ApJ, 489, 1 

Renzini, A., Ciotti, L., D'Ercole, A. \& Pellegrini, S. 1993, ApJ, 419, 52 

Rich, R.M. 1999, in {\em Chemical Evolution from Zero to High Red-shift}, J. Walsh 
\& M. Rosa (eds.), Springer-Verlag, in press 

Ryan, S., Norris, J. \& Beers, T.C. 1996, ApJ, 471, 254  

Searle, L. \& Sargent, W.L.W. 1972, ApJ, 173, 25 

Steidel, C.C., Adelberger, K.L., Giavalisco, M., Dickinson, M. \& Pettini, M. 
1999, ApJ, in press, astro-ph/9811399  
 
Stephens, A. 1999, AJ, 117, 1771 

Thomas, D. 1999, MNRAS, in press, astro-ph/9901226  

Thomas, D., Greggio, L. \& Bender, R.  1999, MNRAS, 302, 537  

Thomas, D. \& Kauffmann, G. 1999, in {\em Spectrophotometric Dating of Stars 
and  Galaxies}, I. Hubeny, S.Heap \& R. Cornett (eds.), ASP Conf. Series, 
astro-ph/9906216   

Tsujimoto, T. \& Shigeyama,  1998, ApJL, 508, L151  

Tsujimoto, T., Shigeyama,  \& Yoshii, Y. 1999, ApJL, in press, 
astro-ph/ 9905057   

Vladilo, G. 1998, ApJ, 493, 583 

Worthey, G. 1994, ApJS, 95, 107  

Worthey, G., Dorman, B. \& Jones, L.A. 1996, AJ, 112, 948  

Worthey, G., Faber, S.M. \& Gonzalez, J.J. 1992, ApJ, 398, 69 
    
} 
\end{document}